\documentclass{article}
\usepackage[utf8]{inputenc}
\usepackage{braket}
\usepackage{amsmath}
\usepackage{graphicx}
\usepackage{enumitem}   
\usepackage{mathrsfs}
\usepackage{dsfont}
\usepackage{mathtools}
\usepackage{amssymb}
\usepackage{xcolor}
\usepackage{physics}
\usepackage{comment}
\usepackage{hyperref}
\usepackage{authblk}

\DeclareMathAlphabet{\pazocal}{OMS}{zplm}{m}{n}

\usepackage[square,numbers]{natbib}

\newcommand{\zerodel}{.\kern-\nulldelimiterspace}

\renewcommand\bra[1]{{\left<#1\right|}}
\renewcommand\ket[1]{{\left|#1\right>}}

\title{Preserving Symmetry: \\ Spontaneous Symmetry Breaking through Decoherence}

\author[1]{Sam Kuypers\thanks{samuelkuypers@gmail.com}}

\affil[1]{Conjecture Institute, United States}

\date{\today}

\begin{document}
\maketitle

\begin{abstract}
Solids appear to have localised centres of mass, yet many-body quantum theory describes them using translationally symmetric models that preclude localisation. Conventionally, this is resolved through spontaneous symmetry breaking by introducing an interaction with a semiclassical environment that breaks the symmetry. In the thermodynamic limit, the interaction can be removed while leaving the state localised. This, however, raises the question of how localisation arises outside the thermodynamic limit, i.e., in finite quantum systems \cite{wallace2018}. Here, we show that, by quantising the environment, the localisation of finite systems occurs within decoherent branches, while the state vector of the composite system remains translationally symmetric. Our approach is analogous to the Page–Wootters construction \cite{page1983evolution} and quantum reference frames; moreover, we recover the semiclassical description as a limiting case while predicting experimentally distinguishable corrections away from this limit.
\end{abstract}

\section{Introduction}\label{sec:one}

Crystals and other solid materials are typically described by translationally invariant models in which the total momentum is conserved. Phases between different total-momentum sectors are inaccessible in the absence of an absolute spatial reference, so the observable properties of such systems are the relational distances and momenta. Accordingly, the system may be taken to occupy a total-momentum eigenstate without affecting any observable predictions. It is therefore striking that solid materials familiar to us from everyday life possess well-localised centres of mass, which necessarily involve superpositions of total-momentum eigenstates.

This problem is well-known from spontaneous symmetry breaking in solids \cite{vanWezel_2019}, where it is conventionally resolved as follows: the models are taken not to be strictly translationally invariant because of couplings to an environment. In standard treatments, the environment is modelled semiclassically, so its effects are represented by an external potential that explicitly breaks the model’s symmetry. The symmetric model is recovered in the limit in which the system size becomes infinite and the symmetry-breaking potential vanishes; in this limit, the Hamiltonian is translationally symmetric, yet the ground state occupies a definite centre of mass due to the system’s sensitivity to the external environment.

Although the standard approach has been operationally successful, these semiclassical models obscure the fact that the system and environment are both quantum and should jointly preserve translation symmetry, which again raises the question of how the system and environment can be in a momentum eigenstate. Similarly, the thermodynamic limit, in which the particle number tends to infinity, raises conceptual issues: in this limit, the Hamiltonian can remain translationally symmetric while admitting a ground state with a definite centre of mass, but actual materials are finite and generally possess unique translationally invariant ground states. 

The present work is an application of the programme by Wallace \cite{wallace2018} to incorporate spontaneous symmetry breaking into decoherence theory, thereby resolving both of the aforementioned problems. As such, we consider a translation-invariant system of \(N\) particles coupled to an environment in a manner that exactly preserves translation symmetry. By analysing the resulting entanglement between the system and environment, we show how decoherence leads to branches of the wavefunction in which the particles' centre of mass is localised and the symmetry appears broken, despite the global state remaining delocalised and translationally symmetric.

This explanation is closely related to what are called \textit{relational approaches} in quantum theory. One of the principal examples is the Page--Wootters construction (see Refs.~\cite{page1983evolution, marletto2017evolution, kuypers2022quantum, rijavec2023robustness, rijavec_conditions_2025}), in which time is treated as an internal degree of freedom of a quantum model, with effective evolution emerging relative to a clock subsystem. Similarly, quantum reference-frame approaches describe physical quantities relative to other quantum systems rather than external backgrounds (see for instance Refs.~\cite{Giacomini2019, Giacomini2021spacetimequantum, Spekkens2007}). The model developed here is similar to those approaches in that it does not posit an absolute spatial background; instead, localisation occurs relative to an environment that functions as a spatial reference.

Finally, the decoherence-based account of spontaneous symmetry breaking presented in this work is not merely conceptually distinct from the conventional one, but the two also differ empirically. For example, as we will show for the case of broken translational symmetry, the decoherence-based account predicts that a solid has a ground-state width that depends not only on the system’s intrinsic properties (as it does in the conventional approach) but also on properties of the environment, most notably its mass. These differences are generally hard to detect because the conventional model of spontaneously broken translational symmetry can be recovered from the decoherence-based one in the limit where the environmental mass is large relative to that of the system. Nonetheless, in principle, they give rise to testable predictions that can distinguish the models, as we will discuss in greater detail in Sec.~\ref{sec:experimental}.

\section{Translational symmetry}
Let us consider a quantum model of \(N\) interacting particles, where the \(i\)-th particle has position and momentum \(X_i\) and \(P_i\) and mass \(m_i\); a general, non-relativistic Hamiltonian of this system is then
\begin{equation} \label{eq:Hamiltonian}
    H = \sum_{j=1}^N \frac{P_j^2}{2m} + V(X_1,\dots,X_N),
\end{equation}
Here, each of the particles' momentum and position is canonically conjugate, and they commute with the position and momentum observables of all other particles so that
\begin{align}
    [ X_i, P_j ] = i \hbar \delta_{ij}, && [P_i, P_j]=[X_i,X_j] = 0 && \forall i,j \in \{1, \dots, N\}.
\end{align}
We shall impose that the Hamiltonian is spatially symmetric, meaning that its potential depends only on the relative positions of the particles. Consequently, the Hamiltonian decomposes into two distinct parts, one describing its centre-of-mass position and momentum, and another describing the relative displacements of the particles. These two components decouple, greatly simplifying the model's behaviour. 

To that end, let us introduce the centre-of-mass position and momentum, as well as the system's total mass, which are defined as follows:
\begin{align}
    P_{\mathrm{CoM}}  := \sum_{j=1}^N P_j, &&   M := \sum_{j=1}^N m_j,
    && X_{\mathrm{CoM}} := \frac{1}{M}\sum_{j=1}^N m_j X_j.
\end{align}
As previously stated, we will assume that the Hamiltonian commutes with \(P_{\mathrm{CoM}}\) so that it is spatially symmetric, i.e. by construction, we impose
\begin{equation}
    [H,P_{\mathrm{CoM}}]=0.
\end{equation}
Due to this symmetry, the Hamiltonian can be expressed as
\begin{equation}
    H = \frac{P_{\mathrm{CoM}}^2}{2M} + H_{\mathrm{int}},
\end{equation}
where \(H_{\mathrm{int}}\) represents the Hamiltonian of the internal degrees of freedom, such as the relative distances of the particles (see Appendix~\ref{app:one} for a full derivation).  

The centre-of- mass position and momentum commute with the relational ones, so that \( X_{\text{CoM}} \) and \( P_{\text{CoM}} \) lie on a Hilbert space that is separate from those of the internal degrees of freedom, implying that the model's Hilbert space factorises as \( \mathcal H = \mathcal H_{\mathrm{CoM}} \otimes \mathcal H_{\mathrm{int}} \). Since superpositions of centre of mass momentum eigenstates are unobservable, the physical states are those for which the centre of mass is entirely spread out. This raises the question of how solids, which are described by \(N\)-body Hamiltonians such as that of Eq.~\eqref{eq:Hamiltonian}, can have localised centres of mass, as they do from experience.

\section{Spontaneous Symmetry Breaking} \label{sec:ssb}

The conventional solution to the aforementioned problem is provided by spontaneous symmetry breaking, and goes as follows: the \(N\) particles are not isolated; they interact with an environment, thereby pinning the crystal down at a particular location. These interactions with the environment are modelled semiclassically, so that the environment is not treated explicitly as a quantum system. Instead, its effects are represented by an external potential that depends on \(X_{\text{CoM}}\), which breaks the model's translational symmetry, thereby localising it. For instance, in Ref.~\cite{vanWezel_2019}, the semiclassical Hamiltonian is 
\begin{equation} \label{eq:semiclassical}
    H_{\text{sc}} (x_0) = \frac{P^2_{\text{CoM}}}{2M} + \frac{\mu}{2} (X_{\text{CoM}}-x_0)^2.
\end{equation}
Here we use that \([X_{\text{CoM}}, P_{\text{CoM}}] = i \hbar\) to recover that this is a harmonic oscillator, whose lowest energy eigenstate is a Gaussian peaked at the position \(x_0\), namely
\begin{align} \label{eq:groundstate}
\ket{\psi_{0}}
=
\left ( \frac{1}{\pi\sigma_{\text{pin}}^{2}} \right )^{1/4} \int^{\infty}_{-\infty} dX
\exp\!\left [
-\frac{(X-x_{0})^{2}}{2\sigma_{\text{pin}}^{2}} 
\right]
\ket{X},
\end{align}
where \(\ket{X}\) is an eigenstate of \(X_{\text{CoM}}\), and where the ground state width is 
\begin{align}
\sigma_{\text{pin}}
:=\left(\frac{\hbar^2}{M\mu}\right)^{1/4}.
\end{align}
The system's translational symmetry is broken by the interactions with its environment since the position \(x_0\) is now singled out as preferred among other positions. This is what allows the lowest energy eigenstate can be localised around \(x_0\) instead of being spread out.

The potential term explains why the system has a particular centre of mass, but at the cost of breaking the model's translation symmetry. However, for an arbitrarily large particle number, the spring constant can be arbitrarily small and still localise the centre of mass, as can be shown by first taking the thermodynamic limit of \(\sigma_{\text{pin}}\) so that \(N\) goes to infinity and subsequently letting \(\mu\) go to zero: \(\lim_{\mu \to 0} \lim_{N \to \infty} \sigma_{\text{pin}} = 0\).\footnote{Note that \(\sigma_{\text{pin}}\)'s dependence on \(N\) is implicit: in a system of \(N\) particles with average mass \(m_{\mathrm{avg}}\), the total mass satisfies \(M = N m_{\mathrm{avg}}\). The thermodynamic limit \(N \to \infty\) therefore corresponds to \(M \to \infty\).} In this limit the Hamiltonian is exactly symmetric, as \(\mu \to 0\) removes the pinning potential, yet its ground state remains perfectly localised at \(x_0\) since \(\sigma_{\text{pin}}\) vanishes. The translation symmetry is said to be broken \textit{spontaneously} because, in the thermodynamic limit, the localised state is an energy eigenstate of the translationally-symmetric Hamiltonian.

But a problem now appears: the limits do not commute since if the system has a finite number of particles, then when \(\mu \to 0\), the centre of mass becomes spread out again even if we take the thermodynamic limit afterward, i.e. \( \lim_{N \to \infty}\lim_{\mu \to 0}  \sigma_{\text{pin}}  = \infty \). This well-known fact \cite{vanWezel_2019} is problematic for an account of spontaneous symmetry breaking in finite systems, as it implies that for finite \(N\), the system's ground state is necessarily translationally invariant when \(\mu \to 0\). Since actual physical systems are decidedly finite, the solution cannot be that symmetry breaking occurs only for systems with infinite particle number.

The model described in Eq.~\eqref{eq:semiclassical} must therefore be interpreted differently. The two successive limits \(\lim_{\mu \to 0} \lim_{N \to \infty} \sigma_{\text{pin}} = 0\) demonstrate that the system is susceptible to environmental decoherence, as arbitrarily small interactions with the environment nonetheless tend to localise the system if it consists of a sufficiently large number of particles. This makes an account in terms of decoherence theory natural but requires that we explicitly model the environment, which we have hitherto neglected to do. 

\subsection{Modelling the environment} \label{sec:modelling-the-environment}

The more complete story is therefore that Eq.~\eqref{eq:semiclassical} describes a Hamiltonian that models the effects of a semiclassical environment. Crucially, if the environment were quantised, the interactions between it and the system would once more preserve translation symmetry, as there would be no fixed spatial reference to embed them into, just as was the case for the initial \(N\)-particle Hamiltonian \cite{MarlettoVedral2022}. The appearance of the broken symmetry can then be explained through decoherence without the need for the thermodynamic limit, a line of thinking that was perhaps first explored by Giulini, Kiefer, and Zeh \cite{Zeh} and more recently, and in greater detail, by Wallace \cite{wallace2018}.

As such, let us explicitly model the environment as a quantum system; in particular, we consider the environment's centre-of-mass degrees of freedom, represented by the canonical position and momentum \(X_0\) and \(P_0\), and by its mass \(M_0\). Let the Hamiltonian for the centre-of-mass degrees of freedom of the \(N\) particles and its coupling to the environment be
\begin{equation} \label{eq:coupled-oscillator}
    H_{\text{SE}} = \frac{P_{\mathrm{CoM}}^2}{2M} +  \frac{P_0^2}{2M_0}  +  \frac{\mu}{2} (X_{\text{CoM}} - X_0)^2.
\end{equation}
Here \( \mu \) is reused, now representing the coupling constant for a coupled harmonic oscillator. For convenience, we omit the internal degrees of freedom. The new total momentum is \( P_{\text{tot}} := P_{\mathrm{CoM}} + P_0 \); we will again restrict attention to a fixed total-momentum sector by assuming that the system is in an eigenstate of \( P_{\text{tot}} \) with eigenvalue zero. 

Generally, this interaction between the two systems will entangle them while preserving \(P_{\text{tot}}\). Such entanglement structure is, for example, evident from the ground state of \(H\), which is
\begin{equation}
    |\Psi_0\rangle \rangle
= \frac{1}{\sqrt{2\pi\hbar}}\left (\frac{1}{\pi\sigma_{\text{rel}}^2} \right )^{1/4}
\int_{-\infty}^{\infty} dX\int_{-\infty}^{\infty} dx\;
\exp\!\Big(-\frac{(X-x)^2}{2\sigma_{\text{rel}}^2}\Big)\,
|X\rangle|x\rangle\;,
\end{equation}
where we use \(\rangle \rangle\) to clarify that this state resides on the Hilbert space of the composite system of the environment and the \(N\) particles, and where the ground-state width of the relative coordinates is
\begin{equation}
\sigma_{\text{rel}} :=\left(\frac{\hbar^2(M+M_0)}{\mu\,MM_0}\right)^{1/4}.
\end{equation}
The absolute state \(|\Psi_0\rangle\rangle\) has two crucial, seemingly contradictory properties that allow us to explain the appearance of symmetry breaking without fundamentally violating translational symmetry. First, note that \(|\Psi_0\rangle \rangle\) is an eigenstate of \(P_{\text{tot}}\) such that the total momentum is zero, as is apparent from the fact that the state is invariant under simultaneous translations of the system and environment \(X_{\text{CoM}} \to X_{\text{CoM}} +a\) and \(X_0 \to X_0 +a\) for arbitrary real \(a\), so that the system is spread out over space. Yet, relative to the environment particle being in state \(\ket{x_0}\), the \(N\)-body system has a well-localised centre of mass, namely
\begin{align}\label{eq:Gaussian}
     \frac{ \langle x_0 |\Psi_0\rangle \rangle } { \| \langle x_0 |\Psi_0\rangle \rangle \| }
= \left(\frac{1}{\pi\sigma_{\text{rel}}^2}\right)^{1/4}
\int_{-\infty}^{\infty} dX\;
\exp\!\left[-\frac{(X-x_0)^2}{2\sigma_{\mathrm{rel}}^2}\right] 
|X\rangle,
\end{align}
where \(\ket{x_0}\) is an eigenstate of \(X_0\) with eigenvalue \(x_0\). The left-hand side of Eq.~\eqref{eq:Gaussian} is called a \textit{relative state} \cite{everett1973theory}, as it describes the centre of mass of degrees of freedom of the \(N\)-body system relative to the environment being at position \(x_0\); the relative state has been normalised, as is evident from the factor \( \| \langle x_0 |\Psi_0\rangle \rangle \|\) in the denominator. A relative state describes what we will call a \textit{branch} of the wavefunction, meaning a part of the wavefunction, often one that evolves approximately autonomously under decoherence \cite{Zurek2003, Zeh}.

The relative state in Eq.~\eqref{eq:Gaussian} is a Gaussian wavepacket, sharply peaked at \(x_0\) if \(\sigma_{\text{rel}}\) is small so that the \(N\)-body system has a well-localised position relative to the environment. Thus, the \(N\) particles can be localised in relative states without the absolute state \(|\Psi_0\rangle \rangle\) breaking translation symmetry. Moreover, there is little to no interference between such relative states on scales of separation much larger than \(\sigma_{\mathrm{rel}}\), as is evident from the reduced density matrix of the \(N\)-particle system, which has the form
\begin{equation}\label{eq:density-matrix}
\int dx_0 \langle x_0 | \Psi_0 \rangle \rangle \langle \langle \Psi_0 | x_0 \rangle
\propto \int dX\,dX'\;
\exp\!\left[-\frac{(X-X')^2}{4\sigma_{\mathrm{rel}}^2}\right]
\,|X\rangle\langle X'|.
\end{equation}
As can be gleaned from Eq.~\eqref{eq:density-matrix}, the density matrix's off-diagonal terms are exponentially suppressed when \(|X' - X | \gg \sigma_{\mathrm{rel}}\). So, on those scales, the reduced density matrix is approximately a classical mixture over different centre-of-mass positions.

Hence, from the perspective of a given branch in which the environment is localised at \(x_0\), the \(N\)-particle system appears to possess a well-defined centre-of-mass position that remains effectively pinned near \(x_0\) with negligible interference from nearby branches. In this way, decoherence accounts for the appearance of a spontaneously broken translational symmetry in branches of the otherwise translationally symmetric wave function, \(| \Psi_0 \rangle \rangle\).

\subsection{Recovering the semiclassical Hamiltonian}\label{sec:decoherence}
The model described in Eq.~\eqref{eq:semiclassical} can be recovered as an effective description of the coupled oscillator of Eq.~\eqref{eq:coupled-oscillator} by assuming that the environment is large relative to the system it couples to. For instance, in the limit in which \(M_0\) goes to infinity, the ground-state width of Eq.~\eqref{eq:coupled-oscillator} reduces to that of a single oscillator in an external potential since
\begin{equation}
\lim_{M_0  \to \infty } \sigma_{\text{rel}} = \sigma_{\text{pin}}.
\end{equation}

The same limit can be used to recover the semiclassical Hamiltonian, which determines the dynamics in branches of the wavefunction. In particular, when \(M_0\) is much larger than \(M\), interference between states with different environment centres-of-mass positions occurs on negligibly long time scales. To demonstrate this, consider some general time-dependent state \(| \Psi (t) \rangle \rangle\) that evolves under \(H\) and is assumed to be an eigenstate of \(P_{\text{tot}}\). As above, we define the state of the \(N\) particles relative to the environment being localised at location \(x_0\) as
\begin{equation}
\ket{\psi(x_0, t)} := \langle x_0|\Psi (t) \rangle\rangle.
\end{equation}
Note that the norm of \( \langle x_0|\Psi(t)\rangle\rangle \) is independent of \(x_0\), as shown in Appendix~\ref{app:relative-state-norm}. This gives us a relative state that describes a particular branch of the wavefunction. We project the dynamics onto this branch by considering \(\ \bra{x_0} H_{\text{SE}} | \Psi (t) \rangle \rangle \) and find that
\begin{equation}\label{eq:branch_term}
i \hbar \partial_t \ket{\psi(x_0,t)}
=
H_{\text{sc}} (x_0)\,\ket{\psi(x_0,t)}
-
\frac{\hbar^2}{2M_0}\,\partial_{x_0}^2 \ket{\psi(x_0,t)} .
\end{equation}
The first term is precisely the symmetry-breaking semiclassical Hamiltonian acting within that branch, while the second term depends on how the global wavefunction varies with the environment coordinate and therefore couples neighbouring branches.

Assume that \( \ket{\psi(x_0,t)}\) varies over a characteristic distance of the ground state \(\sigma_{\text{rel}}\),  which provides a natural lower bound on the position spread since higher-energy eigenstates generally have a larger position spread. Thus, using an order-of-magnitude estimate of each term on the right hand side of Eq.\eqref{eq:branch_term}, we obtain that the effects of the second term in Eq.~\eqref{eq:branch_term} will generally occur on a time scale of order \(\tau_{0} \sim M_0 \sigma_{\text{rel}}^2 /\hbar\), while the time scale on which the semiclassical Hamiltonian operates is of order \(\tau_{\text{eff}} \sim \sqrt{M / \mu}\) so that
\begin{equation}
\frac{\tau_0}{\tau_{\text{eff}}} \sim \sqrt{\frac{M_0(M+M_0)}{M^2}} .
\end{equation}
Consequently, when \(M_0 \gg M\), it holds that \(\tau_0 \gg \tau_{\text{eff}}\), so the effects of the semiclassical Hamiltonian occur on much shorter time scales than the second term of Eq.~\eqref{eq:branch_term}. The semiclassical Hamiltonian is therefore the dominant term, as interference between branches is negligible on time scales of order \(\tau_{\text{eff}}\), so the relative states evolve approximately autonomously under \(H_{\text{sc}} (x_0) )\) on those scales. In this way, the semiclassical Hamiltonian of Eq.~\eqref{eq:semiclassical} is recovered as the approximate Hamiltonian within a branch, even though the full evolution remains translation invariant, as becomes apparent on time-scales of order \(\tau_{0}\) since, on those scales, interference between branches can be observed. This same mechanism can, in principle, explain more general instances of spontaneous symmetry breaking \cite{wallace2018}.

\subsection{Emergence of entanglement} \label{sec:emergence-of-entanglement}
In the preceding section, we studied the ground state of the coupled harmonic oscillator to demonstrate that it exhibits the required entanglement structure. However, the ground state is a stationary state, and as such, the entanglement between the system and the environment is already present in it and does not emerge dynamically. Therefore, we will prove here that a product state can evolve into an entangled one similar to that of Eq.~\eqref{eq:Gaussian}. 

To that end, let us consider again the translation-invariant coupled oscillator model studied earlier in
Eq.~\eqref{eq:coupled-oscillator}. Since the interaction depends only on the relative coordinate \(X_{\mathrm{CoM}}-X_0\), the Hamiltonian can be solved exactly by introducing a total momentum operator and relative coordinates. In particular, define the total momentum as
\begin{equation}
    P_{\mathrm{tot}}
    :=
    P_{\mathrm{CoM}}+P_0,
\end{equation}
and the relative momentum
\begin{equation}
    p_{\mathrm{rel}}
    :=
    m_{\text{red}}\left(
    \frac{P_{\mathrm{CoM}}}{M}
    -
    \frac{P_0}{M_0}
    \right),
\end{equation}
with
\begin{equation}
    m_{\text{red}} := \frac{MM_0}{M+M_0},
\end{equation}
being the reduced mass. In terms of these variables, the Hamiltonian becomes
\begin{equation}
    H_{\text{SE}}
    =
    \frac{P_{\mathrm{tot}}^2}{2(M+M_0)}
    +
    \frac{p_{\mathrm{rel}}^2}{2m_{\text{red}}}
    +
    \frac{1}{2}m_{\text{red}}\omega^2
    (X_{\mathrm{CoM}}-X_0)^2,
\end{equation}
where \(\omega:=\sqrt{\mu/m_{\text{red}}}\). The dynamics, therefore, separate into a free
centre-of-mass contribution and a Hamiltonian for the relative position and momentum that corresponds to a harmonic oscillator.

Suppose that the initial state is unentangled, in that it is a state of momentum eigenstates of both the system and environment:
\begin{equation} \label{eq:Product-state}
    |\Psi(0)\rangle \rangle
    =
    |p\rangle_{\text{CoM}}
    |-p\rangle_0.
\end{equation}
The initial state is delocalised in two respects. First, it is an eigenstate of  \(P_{\text{tot}}\) with eigenvalue \(0\) and therefore exhibits no preferred location in absolute space. Second, each subsystem is itself in a momentum eigenstate, so neither possesses a definite position relative to the other. Consequently, the initial state contains no branches in which the system occupies a well-defined location relative to the environment.

Despite the initial state being unentangled, the system-environment interaction entangles them, creating branches in which the system has a well-defined position relative to the environment. The total momentum once more decouples from the internal degrees of freedom, and the internal degrees of freedom, described by \(p_{\text{rel}}\) and \(X_{\text{CoM}}- X_0\), have a Hamiltonian that is equivalent to that of a harmonic oscillator. As we prove in Appendix~\ref{app:one}, the initial state shown in Eq.~\eqref{eq:Product-state} becomes an eigenstate of \(X_{\text{CoM}} - X_0\) at \(t_0= \pi / 2\omega\), namely
\begin{equation} \label{eq:exact-displacement}
    |\Psi(t_0)\rangle \rangle
    \propto
    \int dX\;
    \left| X+ \frac{p}{m_{\text{red}}\omega} \right\rangle_{\mathrm{CoM}}
    |X\rangle_0.
\end{equation}
Equation~\eqref{eq:exact-displacement} shows that the state at \(t_0\) consists of branches in which the system and environment possess a definite relative displacement. Although both \(X_0\) and \(X_{\text{CoM}}\) are non-sharp at \(t_0\), the relative position \(X_{\mathrm{CoM}}-X_0\) is sharp at \(t_0\) because there is a definite displacement of \(p/m_{\text{red}}\omega\) between the two systems' positions. The interaction has therefore generated branches in which the system is localised relative to the environment. Notably, because the harmonic oscillator undergoes periodic motion, this process is reversible: the entanglement later disappears, and the branches vanish with it.

\section{Internal reference frames}
The preceding discussion suggests a close analogy with the Page–Wootters construction (see e.g. Refs.~\cite{page1983evolution, marletto2017evolution, kuypers2022quantum, rijavec2023robustness, rijavec_conditions_2025}). In that construction, one imposes the constraint that the system is in an energy eigenstate, which is a stationary state so that the classical time parameter becomes unobservable. One can then recover the Schrödinger equation in the Page–Wootters model by conditioning on the state of a \textit{clock}, which is an isolated system that is entangled with the rest of the universe. The emergence of time is then explained relationally relative to the reference provided by the clock; this is also closely aligned with quantum reference-frame approaches, in which physical quantities are defined relative to quantum systems rather than classical coordinates (see Refs.~\cite{Giacomini2019, Giacomini2021spacetimequantum, Spekkens2007}). 

In this work, spatial translations play the corresponding role: consider the centre-of-mass degree of freedom of the system together with a reference particle. Defining the total momentum as
\begin{equation}
    P_{\mathrm{tot}} := P_{\mathrm{CoM}} + P_0,
\end{equation}
and imposing that
\begin{equation} \label{eq:momentum-constraint}
    P_{\mathrm{tot}}|\Psi\rangle\rangle = 0,
\end{equation}
removes any absolute notion of location at the level of the composite state. But given a relative state
\begin{equation}
    |\psi(x)\rangle := \langle x|\Psi\rangle\rangle,
\end{equation}
we can project onto the constraint of Eq.~\eqref{eq:momentum-constraint} to obtain
\begin{equation} \label{eq:P-COM}
    i\hbar\,\partial_x |\psi(x)\rangle
    = P_{\mathrm{CoM}} |\psi(x)\rangle ,
\end{equation}
Notably, the variable \(x\) is an internal degree of freedom of the model, representing the location of an environment. The corresponding global state which satisfies Eq.~\eqref{eq:P-COM} is
\begin{equation}
    |\Psi\rangle\rangle
    = \int dx\;
    e^{-\frac{i}{\hbar}xP_{\mathrm{CoM}}}|\psi(0)\rangle |x\rangle .
\end{equation}
When the reference system is sufficiently massive, distinct branches evolve approximately independently, allowing the environment to function as an internal spatial frame.

Finally, the relational approach also admits a Heisenberg-picture-like counterpart, in which the spatial degree of freedom is moved to the operators via a unitary transformation. This allows observables to be defined as functions of the position operator \(X_0\), with derivatives given by commutators with \(P_0\). As discussed in Appendix~\ref{app:heisenberg}, this defines derivatives entirely in terms of physical observables.

\section{Experimental signatures} \label{sec:experimental}
The discussion so far has been largely theoretical, but the decoherence-based theory of spontaneous symmetry breaking developed here is empirically distinct from the conventional semiclassical account. Consider, for instance, that in conventional models, such as that of Eq.~\eqref{eq:semiclassical}, the ground-state width \(\sigma_{\text{pin}}\) is independent of the environmental mass \(M_0\), since the environment does not explicitly appear in the semiclassical treatment. Our model predicts, instead, that the width does depend on the environmental mass, as demonstrated for \(\sigma_{\text{rel}}\) in Sec.~\ref{sec:modelling-the-environment}, making the two theories empirically distinguishable in principle. Although this dependence is small when \(M_0 \gg M\), the two quantities differ significantly when \(M_0\) is comparable in size to \(M\).

A second difference is that the semiclassical model contains no mechanism by which distinct symmetry-broken branches can interfere. In the present theory, interference between these branches may become observable on sufficiently long time scales, as demonstrated in Sec.~\ref{sec:emergence-of-entanglement}. Such interference effects are unlikely to be experimentally accessible in macroscopic systems. However, in mesoscopic systems interacting with an environment of comparable mass, they may become observable directly or indirectly by giving rise to other observable signatures.

These effects are subtle but might be detectable in controlled settings in which a material is coupled to an artificial environment. One possible setting is a controllable quantum system in which a mesoscopic object interacts with a mechanical environment. For instance, in optomechanical systems, the interaction strength, among other properties of the effective environment, can be controlled~\cite{Aspelmeyer2014}. Such systems may therefore provide a means of testing the predicted differences between the physical properties of the system in the present theory and those predicted by the conventional semiclassical description, allowing us to distinguish the two approaches. 

Moreover, similar considerations should, in principle, apply to other forms of spontaneous symmetry breaking, such as the spontaneous breaking of rotational symmetry in ferromagnets. Such systems may make testing the corresponding predictions of the decoherence-based approach more feasible than it is for translational symmetry, although the underlying mechanism is the same. These examples demonstrate the existence of empirical differences and suggest possible methods by which they could be observed experimentally; the development of a detailed experimental test is left to future research.

\section{Conclusion}
Building on the work of Wallace \cite{wallace2018}, we have shown that the apparent localisation of the centre of mass does not require any fundamental breaking of translation symmetry. Instead, the symmetry can appear broken only relative to an environment. By modelling the environment explicitly and analysing the resulting entanglement structure, we find that symmetry-broken behaviour emerges only within relative states, while the global state remains delocalised and translationally symmetric. Due to decoherence, the branches in which the symmetry appears broken do not interact significantly with one another, thereby hiding the symmetry from view.

The decoherence-based account of spontaneous symmetry breaking does not require the system to be initially entangled since such entanglement can arise dynamically through interaction with the environment; this is expected to be a generic feature of such models, much as decoherence itself is. Moreover, the approach recovers the semiclassical model in the limit of an infinitely massive environment.

Because the decoherence-based account assumes a quantum environment (whereas the semiclassical approach by definition does not), the two approaches produce different predictions. For example, it predicts corrections to the ground-state width of a system’s position that depend on the environmental mass. These predictions make the approaches empirically distinguishable in principle and may allow a critical test of whether the environment should be quantised.

\section*{Acknowledgements}
The author wishes to thank Chiara Marletto, Charles Alexandre Bédard, David Deutsch, and Eric Marcus for their valuable feedback. This work was funded by Conjecture Institute.

\appendix

\section{Derivation of the Centre-of-Mass Hamiltonian}\label{app:one}
Here we derive the standard result that a translationally symmetric Hamiltonian decomposes into a centre-of-mass component and one that governs the internal degrees of freedom of the system. Consider a non-relativistic system of \(N\) particles with Hamiltonian
\begin{equation}
H=\sum_{j=1}^N\frac{P_j^2}{2m_j}+V(X_1,\dots,X_N).
\end{equation}
We impose that the system is translationally symmetric, i.e.
\begin{equation}
[H,P_{\mathrm{CoM}}]=0,
\end{equation}
and define the centre-of-mass and relative coordinates as
\begin{align}
X_{\mathrm{CoM}}&:=\frac{1}{M}\sum_{j=1}^N m_jX_j,\\
P_{\mathrm{CoM}}&:=\sum_{j=1}^N P_j,\\
x_j&:=X_j-X_{\mathrm{CoM}},\\
\pi_j&:=P_j-\frac{m_j}{M}P_{\mathrm{CoM}}.
\end{align}
The relative coordinates satisfy
\(
\sum_{j=1}^N x_j=\sum_{j=1}^N\pi_j=0,
\)
so only \(N-1\) of them are independent. Moreover, their commutation relations are
\begin{equation}
[X_{\mathrm{CoM}},P_{\mathrm{CoM}}]=i\hbar,
\end{equation}
and
\begin{equation}
[X_{\mathrm{CoM}},x_j]
=[X_{\mathrm{CoM}},\pi_j]
=[P_{\mathrm{CoM}},x_j]
=[P_{\mathrm{CoM}},\pi_j]
=0.
\end{equation}
Translation invariance implies that the potential depends only on the relative coordinates, so we may define
\begin{equation}
V_{\mathrm{rel}}(x_1,\dots,x_N)
:=
V(X_1,\dots,X_N).
\end{equation}
Although \(V_{\mathrm{rel}}\) is here expressed as a function of \(N\) variables, only \(N-1\) of them are independent because of the above constraint on the relative coordinates.

The Hamiltonian can now be shown to decompose into a term for the centre-of-mass degrees of freedom and one for the internal degrees of freedom:
\begin{equation}
H
=
\frac{P_{\mathrm{CoM}}^2}{2M}
+
H_{\mathrm{int}},
\end{equation}
\label{eq:int-ham}
where the Hamiltonian for the internal degrees of freedom is
\begin{equation}
H_{\mathrm{int}}
:=
\sum_{j=1}^{N}
\frac{\pi_j^2}{2m_j}
+
V_{\mathrm{rel}}(x_1,\dots,x_N).
\end{equation}
Thus, as is evident from Eq.~\eqref{eq:int-ham}, the Hamiltonian \(H_{\mathrm{int}}\) depends only on the internal degrees of freedom. It can be readily verified that \(H_{\mathrm{int}}\) commutes with both \(P_{\mathrm{CoM}}\) and \(X_{\mathrm{CoM}}\), so that the internal and centre-of-mass degrees of freedom effectively represent different, non-interacting systems.

\section{Entanglement} \label{app:two}
The relative degrees of freedom evolve as a harmonic oscillator, 
\begin{equation}
H_{\mathrm{rel}}
=
\frac{p_{\mathrm{rel}}^2}{2m_{\mathrm{red}}}
+
\frac{1}{2}m_{\mathrm{red}}\omega^2x_{\mathrm{rel}}^2,
\end{equation}
where \(x_{\mathrm{rel}}:=X_{\mathrm{CoM}}-X_0\). We now use the well-known result that, in the Heisenberg picture, the observables of the harmonic oscillator have the following time-dependent expression
\begin{equation}
x_{\mathrm{rel}}(t)
=
x_{\mathrm{rel}}(0)\cos(\omega t)
+
\frac{p_{\mathrm{rel}}(0)}{m_{\mathrm{red}}\omega}\sin(\omega t).
\end{equation}
Consequently, at \(t_0= \pi / 2 \omega\), we have
\begin{equation}
x_{\mathrm{rel}}(t_0)
=
\frac{p_{\mathrm{rel}}(0)}{m_{\mathrm{red}}\omega}.
\end{equation}
Since the initial state of Eq.~\eqref{eq:Product-state} is an eigenstate of \(p_{\mathrm{rel}}\) with eigenvalue \(p\), it follows that the evolved state at \(t_0\) is an eigenstate of \(X_{\mathrm{CoM}}-X_0\) with eigenvalue \(p/(m_{\mathrm{red}}\omega)\). It can also readily be shown that the variances of \(X_{\text{CoM}}\) and \(X_0\) are infinite throughout. Therefore, the state of the system in the Schrödinger picture must be
\begin{equation}
|\Psi(t_0)\rangle\rangle
\propto
\int dX\;
\left|X+\frac{p}{m_{\mathrm{red}}\omega}\right\rangle_{\mathrm{CoM}}
|X\rangle_0,
\end{equation}
which is precisely Eq.~\eqref{eq:exact-displacement}.

\section{Constant norm}
\label{app:relative-state-norm}

We shall prove that the norm of a relative state such as
\begin{equation}
|\phi(x,t)\rangle := \langle x|\Psi(t)\rangle\rangle .
\end{equation}
is constant in \(x\) if \(P_{\mathrm{tot}}|\Psi(t)\rangle\rangle=0\).

First, we project the constraint \(P_{\mathrm{tot}}|\Psi(t)\rangle\rangle=0\) onto \(\langle x|\), which gives us
\begin{equation}
i\hbar\,\partial_x|\phi(x,t)\rangle
=
P_{\mathrm{CoM}}|\phi(x,t)\rangle .
\end{equation}
It follows that
\begin{align}
\partial_x\langle\phi(x,t)|\phi(x,t)\rangle
&=
\frac{i}{\hbar}
\langle\phi(x,t)|P_{\mathrm{CoM}}|\phi(x,t)\rangle
-
\frac{i}{\hbar}
\langle\phi(x,t)|P_{\mathrm{CoM}}|\phi(x,t)\rangle \\
&=0 .
\end{align}
Evidently, \( \|\langle x|\Psi(t)\rangle\rangle\| \) is independent of \(x\).

\section{Heisenberg-picture-like construction} \label{app:heisenberg}

The relational formulation developed in this work places the spatial reference in the state vector. Part of this relational structure may be moved to the observables instead \cite{kuypers2022quantum}, resulting in a construction that is similar to the Heisenberg picture. Unlike the conventional Heisenberg picture, however, the transformation is not generated by the Hamiltonian and does not eliminate the time dependence of the state. Instead, it transfers the spatial reference from the state to the observables. The benefit is that this allows differentiation to be formulated entirely in terms of physical observables rather than with respect to an external classical parameter. 

To that end, let
\begin{equation}
    U
    :=
    e^{\frac{i}{\hbar}X_0P_{\mathrm{CoM}}}.
\end{equation}
Given a Schrödinger-picture observable \(A\) and state \(\ket{\Psi}\), we define
\begin{align}
    A
    &\longrightarrow
    A(X_0)
    :=
    UAU^\dagger,
    \\
    \ket{\Psi(t)}
    &\longrightarrow
    | \widetilde{\Psi}(t) \rangle :=  U\ket{\Psi(t)}.
\end{align}
For example, the transformed total momentum is
\begin{align}
    UP_{\mathrm{tot}}U^\dagger
    \nonumber
    =
    P_0.
\end{align}
Similarly, the transformed relative-position observable is
\begin{equation}
    U(X_{\mathrm{CoM}}-X_0)U^\dagger
    =
    X_{\mathrm{CoM}}.
\end{equation}
Thus, the relational position is represented by \(X_{\mathrm{CoM}}\) in the transformed picture, while the total-momentum constraint is transferred to the environment.

The observable, when evaluated at a specific environmental position \(x\), may be expressed as
\begin{equation}
    A(x)
    :=
    e^{\frac{i}{\hbar}xP_{\mathrm{CoM}}}
    A
    e^{-\frac{i}{\hbar}xP_{\mathrm{CoM}}},
\end{equation}
so that, using the position eigenstates of the environment \(\ket{x}\), we find that
\begin{equation}
    A(X_0)\ket{x}
    =
    A(x)\ket{x}.
\end{equation}
We can define a natural notion of differentiation relative to the observable \(X_0\), despite that observable remaining fixed, namely
\begin{equation}
    \frac{dA(X_0)}{dX_0}
    :=
    \frac{i}{\hbar}[P_0,A(X_0)],
\end{equation}
which reproduces the ordinary derivative when acting on an eigenstate of \(X_0\), that is
\begin{equation}
    \frac{dA(X_0)}{dX_0}\ket{x}
    =
    \frac{dA(x)}{dx}\ket{x}.
\end{equation}

Finally, the transformation preserves the total-momentum constraint: since \(\ket{\Psi(t)}\) is an eigenstate of \(P_{\mathrm{tot}}\), it straightforwardly follows that
\begin{equation}
    UP_{\mathrm{tot}}U^\dagger | \widetilde{\Psi} \rangle
    =
    0.
\end{equation}
Thus, the transformed state remains an eigenstate of the transformed total-momentum operator.
 
\bibliographystyle{unsrtnat}
\bibliography{mybib}

\end{document}